\documentclass[usenatbib]{mn2e}
\usepackage{graphics,longtable}
\usepackage{graphicx}
\usepackage{float}
\usepackage{txfonts}
\usepackage{array}
\usepackage{multirow}
\usepackage{color}
\usepackage{mathptmx}

\title[Measuring the time delays in the presence of microlensing]
{A simple method to determine the time delays in presence of
microlensing: application to HE 0435-1223 and PG1115+080}
\author [Tsvetkova V.S. et al.] { \parbox[t]{\textwidth}{
\vspace{-1.0cm}
      V.S.Tsvetkova$^{1,2}$ \thanks {E-mail: tsvet999@gmail.com},
       V.M.Shulga$^{1,3}$,  L.A.Berdina$^1$ }
       \vspace*{6pt}\\
       $^1$Institute of Radio Astronomy of Nat.Ac.Sci. of Ukraine,
           Krasnoznamennaya 4, 61002 Kharkov, Ukraine\\
       $^2$Institute of Astronomy of Kharkov National University, Sumskaya
           35, 61022 Kharkov, Ukraine\\
       $^3$V.N.Karazin Kharkov National University, Svobody sq. 4, 61070
       Kharkov, Ukraine
       \vspace*{-0.5cm}}

\pubyear{...}

\begin{document}
\label{firstpage}
\pagerange{\pageref{firstpage}--\pageref{lastpage}}
\maketitle

\begin{abstract}
A method for determining the time delays in gravitationally lensed quasars is
proposed, which offers a simple and transparent procedure to
mitigate the effects of microlensing. The method is based on fundamental
properties of representation of quadratically integrable functions by their
expansions in orthogonal polynomials series. The method was tested on the
artificial light curves simulated for the Time Delay Challenge campaign TDC0.
The new estimates of the time delays in the gravitationally
lensed quasars HE 0435-1223 and PG 1115+080 are obtained and compared with
the results reported by other authors earlier.
\end{abstract}


\section{Introduction}

One of the potential astrophysical applications of measuring the
differential time delays in gravitationally lensed quasars is a
possibility to determine the Hubble constant value with no need
of the intermediate standard candles. This was
noticed for the first time by S.Refsdal (1964) long before the
discovery of the first gravitationally lensed quasar, QSO 0957+561
(Walsh, Carswell \& Weymann 1979). Knowledge of the time delays is
necessary for many other applications in astrophysics and cosmology,
such as studies of matter distribution at different spatial scales
in the Universe, including the dark matter, investigation of spatial
structures of lensing galaxies, etc.

The first attempt to measure the time delay has been made for QSO 0957+561
(Florentin-Nielsen 1984). Multiple further attempts provided noticeably
diverging values for the time delay in this doubly imaged quasar (Schild
\& Cholfin 1986; Vanderriest et al. 1989; Press, Rybicki \& Hewitt 1992),
thus demonstrating complexity of the problem. This is due to a number of
objective factors, such as: small amplitudes of the quasar intrinsic
brightness variations, which are often comparable with the photometry
errors; presence of microlensing events; the random flux transfer between
the components in photometry of blended images; difficulties in providing
the long-term uninterrupted monitoring with a sufficient sampling rate and
high photometric precision. The detailed list of these factors can be found,
e.g., in Tewes, Courbin \& Meylan (2013). As a result, a consensus about
the time delay value in the lens QSO 0957+561 was attained as late
as in 1997 (Kundi\v c et al. 1997; Schild \& Thompson 1997).

During 1980-1992, fundamentals of determining the time delays in
gravitationally lensed quasars has been elaborated (Press et al. 1992;
Pelt et al. 1994). In the next years several methods to measure the
time delays have been proposed based, in one form or another, on the
approach developed in these pioneer works, (Schechter et al. 1997;
Barkana 1997; Burud et al. 2000; Kochanek et al. 2006; Eulaers \&
Magain 2011; Courbin et al. 2011; Tewes et al. 2012, and other).
In recent years, an increasing interest to the problem of measuring
the time delays is observed from the astronomical community. To a
considerable degree, this is caused by expectations of a huge data
flow on the newly discovered strong lenses from the Dark Energy Survey,
PanSTARRS, LSST and other survey programs when they  become operational.
In spite of a possible bias in the determination of the Hubble constant
from the time delay technique due to the mass-sheet degeneracy (Falco
et al. 1985; Xu et al. 2015), it is believed to be an important tool in
cosmological studies.

Estimation of the Hubble constant from the time delays requires them
to be measured with a rather high precision: according to Kochanek
\& Schechter (2004) the relative error of the order of 1\% is needed.
Until recently, the precision of time delay measurements
was as a rule much lower. During the last few years, the situation is
tending clearly to change. Thanks to the efforts of some targeted programs
(e.g. COSMOGRAIL), the well-sampled light curves of a long duration
and with rather short gaps between the seasons appeared for some objects
and became publicly accessible, (e.g. Courbin et al. 2011; Eulaers et al.
2013; Rathna Kumar et al. 2013; Tewes et al. 2013). This has given rise
to creation of new methods and versions of the already existing ones.
Liao et al. (2015) report about seven teams participating in a blind
signal processing competition named Time Delay Challenge 1 (TDC1), who
submitted results from 78 different methods. They note that in processing
their mock light curves, several methods have given the accuracy of $\leq0.03$,
while some of the methods have already reached sub-percent accuracy.

A family of methods known as the point estimators does not provide in
a direct and explicit form the estimates of errors of determining the
time delays, which would be an immediate output of processing the
light curves. To obtain the time delay error estimate, an additional
procedure is usually used, known as Monte Carlo simulation. In some
works, e.g., Morgan et al. (2008), Hojjati et al.(2014),
a statistical approach based on Gaussian process modeling is used.
The approach does not need Monte Karlo simulation to estimate the
error of the time delay determination and provides its own estimate
of the uncertainty as a natural result of the whole procedure.
\begin{figure*}
\resizebox{17cm}{!}{\includegraphics{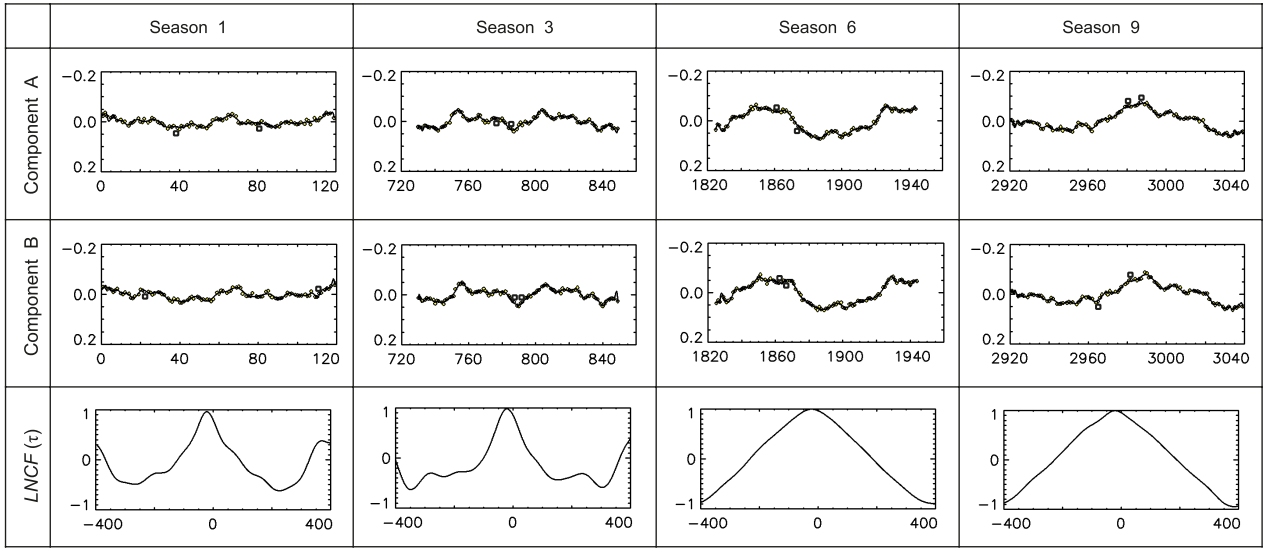}}
 \caption {Synthetic TDC0 light curves "rung2-pair4"
 and their regressions by series expansion in Legendre polynomials of
 the 39-th order (two  upper rows), and the corresponding cross-correlation
 functions  LNCF($\tau$) calculated  according to Eq. 3. The data points,
 which  are more than 3$\delta_{appr}$ offset from the regression curve are
 indicated by the open circles. }
\end{figure*}

Most of methods needs some algorithms to build a model light curve. In
doing so, a necessity arises to properly interpolate the unevenly sampled
data points in the light curves under consideration, and this is one of
the main technical problems in determining the time delays. A variety of
interpolating functions and algorithms is used, such as polynomial
approximation (Lehar et al. 1992; Kochanek et al. 2006), spline interpolation
(Tewes et al. 2013b; Barkana 1997), smoothing with the sampling function
(Vakulik et al. 2009) or with a linear combination of Gaussian kernels,
(Cuevas-Tello, Tino \& Raychaudhury 2006).

All these approaches, while differing in algorithms of the initial data
interpolation, use, in one form or another, the cross-correlation maximum
or mutual dispersion minimum criteria to find the time delay estimate.
In some cases, the light curves of the lensed components are analysed in
pairs, while sometimes, for example, for multiply imaged systems, the
values of the time delays are determined from a joint analysis of light
curves of all image components.

One of the most serious complications in time delay determination is due
to microlensing events, which distort the intrinsic quasar light curves
differently in different quasar images. The choice of a method to
eliminate the effect of microlensing in each specific case
depends strongly on characteristics of the quasar intrinsic
variability and the variability caused by microlensing, in particular,
on relationship between the typical amplitudes and time scales of both
processes. The work by the participants of the
COSMOGRAIL project (Tewes, Courbin \& Meylan 2013) is dedicated to
elaboration of methods for determining the time delays in presence of
"slow" microlensing events, that is, the events with the characteristic
time scale of variability exceeding that
of the quasar variability.

\section {The proposed method }
Our approach to determine the time delays implies a pair-wise
comparison of light curves, with both of them represented by their
polynomial regressions. To approximate the initial light curves,
we use their representations as the series expansions in orthogonal
polynomials.

\subsection {Regression procedure}
Approximations and series expansions in normalized orthogonal functions
are known to possess a number of useful properties, (Korn \& Korn 2000,
Secs.15.2-6, 20.6-2), which provide certain convenience and flexibility
in practice. In particular, approximation of an arbitrary quadratically integrable function $f(x)$ by an
orthonormal set, say, $u_0(x), \, u_1(x), \, u_2(x), \, ... ,\, u_n(x)$ of the
form $f(x)\approx s_n(x)=a_0u_0(x)+a_1u_1(x)+a_2u_2(x)+ ... +a_nu_n(x)$
has the advantage that exclusion of some terms from the sum $s_n(x)$ or
addition of a new extra term $a_{n+1}u_{n+1}(x)$ leave
the values of previously computed coefficients $a_0, a_1, a_2,..., a_n$
unchanged.

\begin{figure*}
\resizebox{17cm}{!}{\includegraphics{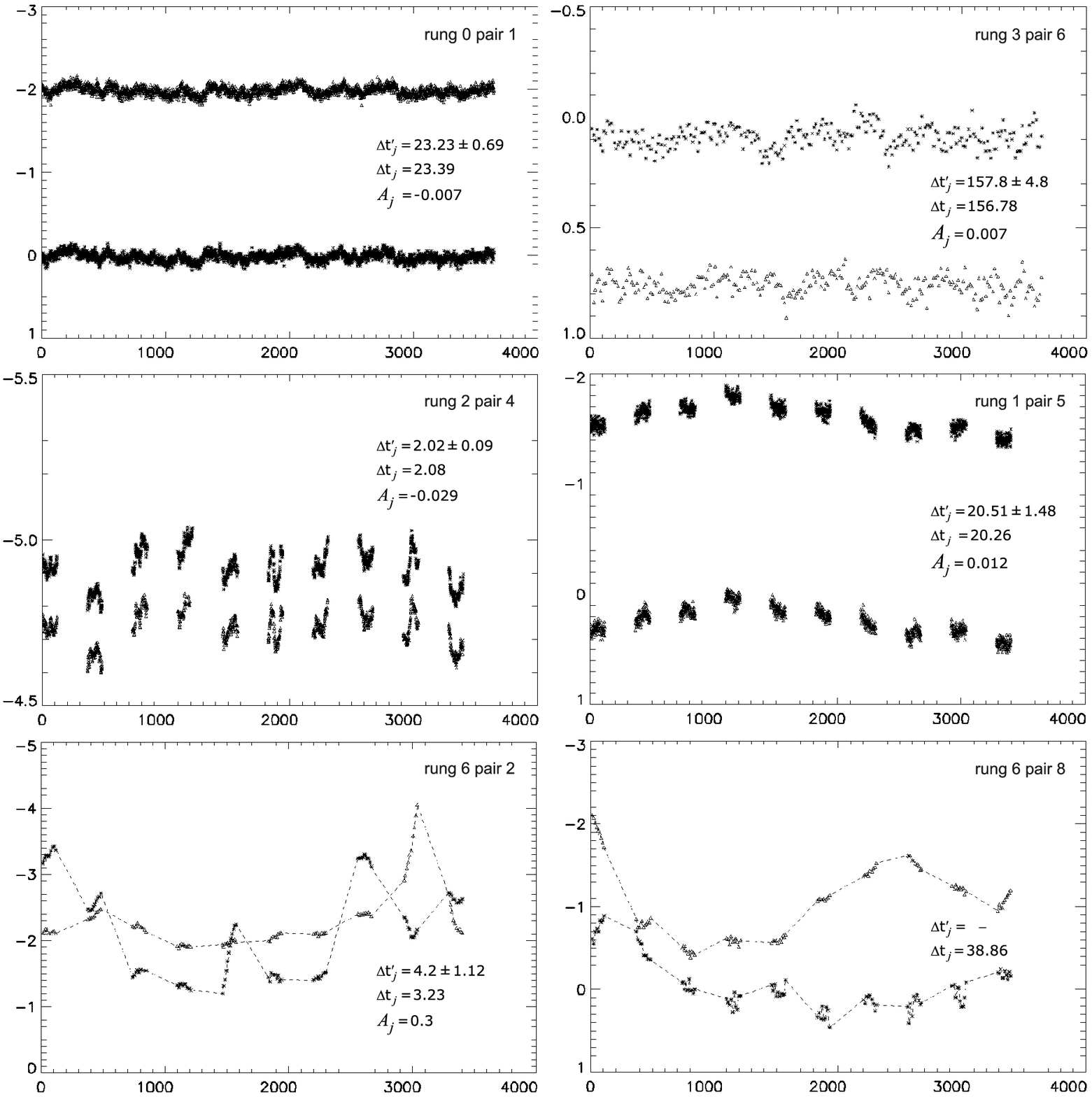}}
\caption{Some results of participation in a Time Delay Challenge (TDC0)
campaign. Six pairs of the simulated light curves of various types are
presented. The quantities in the panels are: the time delay estimates
$\Delta t^\prime_j$ determined with the use of the proposed method and
the corresponding uncertainties $\delta_j$ calculated as described in
Sec. 2; the true time delays $\Delta t_j$, and the relative deviations
of particular estimates from the corresponding true values, $A_j={{\Delta
t^\prime_j-\Delta t_j}\over \Delta t_j}$, -- "accuracy" or "bias" of an
individual determination according to Liao et al. (2015). }
\end{figure*}

In the works of some authors, for example, Kochanek et al. (2006),
Lehar et al. (1992), Legendre polynomials are used, which are orthogonal
only at the continuous set of points specified in a finite interval with
a constant weighting function. To make use of the advantages of the
orthogonal-polynomial regressions, we must start from
constructing an
orthonormal basis specified at a discrete set of unevenly spaced data
points (the dates of observations in our case). Such a basis can be
constructed through the Gram-Schmidt orthogonalization procedure (Korn
\& Korn 2000, 15.2-5). Any arbitrary set that represents a complete
function system specified at the dates of observations can be accepted
in this procedure as an initial basis. In our case, the system of
Legendre polynomials turned out to be the best one in the sense of
computational stability.

In a general case, if $f(x)$ is given at a discrete set of $m+1$ points
$x_0,\, x_1, ...,\, x_m$, the coefficients $a_i$ in $s_n(x)$ are determined
(Korn \& Korn 2000, 20.6-2) by the expression:
\begin{equation}
a_i={{\sum_{k=0}^m \gamma_k f(x_k)u_i(x_k)}\over{\sum_{k=0}^m
\gamma_ku_i^2(x_k)}}
\end{equation}
where $\gamma_k$ are the weights, which are related to photometric
uncertainties and often put to equal unity, and $i=0,1,2,...,n,
(n \leq m)$.

The property indicated above provides a very simple procedure to exempt
the light curves of the lensed quasar components from the effects of
microlensing events, at least for those lenses where the microlensing
brightness variations are much slower as compared to the quasar intrinsic
brightness fluctuations. Making use of the properties of the
orthonormalized polynomials noted above, we may exclude for such
lenses the lower-order terms from a series approximating the light curve
of a certain image component without a necessity to recalculate
coefficients of the polynomial regression. In doing so, we remove the
linear (or quadratic if needed) trends inherent both in microlensing
and in the intrinsic variability of a quasar. Besides, we obtain the
light curve representations all reduced to the zero average level.
Naturally, we may return these lower-order terms, thus recovering the
average level for each component and, if needed, using them to represent
the microlensing light curves without any additional calculations.

There may be the data points in the observed light curves, which are
offset from the regression curve by a quantity exceeding noticeably
both the RMS error of photometry $\delta_{phot}$ and the approximation
error, $\delta_{appr}$, which is the RMS deviation of the data points
from the polynomial regression calculated and displayed in running the
program. We permitted that up to two or three data points, which are
more than 3$\delta_{appr}$ offset from the regression curve, be
identified by the program. Then these points get the values inherent
in the regression curve in their locations, and the approximation
routine is repeated with the data points modified in this way.

To select the maximal "reasonable" order of the polynomial regression,
we were guided, on one hand, by a behaviour of the RMS errors of
approximation in its dependence on the polynomial order. In particular,
the error of approximation $\delta_{appr}$ must be close to the error
typical of the initial photometry data, $\delta_{phot}$, but must not
exceed it: $\delta_{appr}\leq\delta_{phot}$. On the other hand,
oscillations in the regression curves emerging sometimes at the
ends of observational seasons, must not exceed the photometry errors
in amplitudes. This may serve as a constraint on the upper limit of
the polynomial order.

\subsection {Calculation of cross-correlations and uncertainties}

The further analysis consists in calculation of the cross-correlation
function CF$(\tau)$ for a corresponding pair of light curves
represented by the values of their polynomial regressions $f(t_k)$
and $g(t_k)$ in the evenly sampled data points:
\begin{equation}
CF(\tau)={1\over M}\sum_{k}{{[f(t_k)-f^*][g(t_k+\tau)-g^*]}
\over{\sqrt {D_f D_g}}}
\end{equation}

Here, $f(t_k)$ and $g(t_k)$ are the values of the approximating
polynomials in the corresponding points, $f^*$ and $g^*$ are their
mean values at the considered interval, $\tau$ is the time lag,
$M$ is a number of common points in $f(t_k)$ and $g(t_k)$, which
participate in calculations of $CF(\tau)$, and $D_f$, $D_g$ are
variances of $f(t_k)$ and $g(t_k)$.

As one data set slides across the other in calculating the
cross-correlation function, the data points near the signal edges
fall out of the calculations successively, thus resulting in
distortion of the cross-correlation function. To exclude such edge
effects, we used a cross-correlation procedure similar to the
Locally Normalized Discrete Correlation Function, (LNDCF), proposed
by Lehar et al. (1992). Namely, we replaced the mean values $f^*$
and $g^*$ and their variances $D_f$ and $D_g$ with their current
values $f^*_\tau$ and $g^*_\tau$, $D_{f\tau}$ and $D_{g\tau}$
corresponding to the given time lags $\tau$:
\begin{equation}
LNCF(\tau)={1\over M_\tau}\sum_{k}{{[f(t_k)-f^*_\tau][g(t_k+\tau)
-g^*_\tau]}\over{\sqrt{D_{f\tau} D_{g_\tau}}}}
\end{equation}

For this Locally Normalized Correlation Function, LNCF$(\tau)$, the
program is searching for the maximum, and its position is then accepted
as an estimate of the time delay for a particular image pair.

In estimating the uncertainties $\delta_j$ of our time delay
measurements, we do not use the Monte Carlo simulation, but proceed
in the following way. The estimates of statistical parameters of a
stationary random process are known to be equally valid both from
the analysis of the whole signal record, and by averaging the results
of processing separate realizations (subsets) of the process.
Therefore, we accept the estimates of the time
delays for individual seasons averaged over the seasons as the most
probable values of the time delays. The RMS deviation of the time
delays for separate seasons from their average over the seasons is
proposed to be treated as a measure of uncertainty $\delta_j$ of a
particular time delay determination.

In Fig. 1 we show some examples of the synthetic light
curves proposed to the astronomical community by the Time Delay Challenge
(TDC) campaign (Liao Kai et al. 2015). The light curves "rung2-pair4"
of the TDC0 issue and their regressions by expansions in Legendre
polynomial series calculated in the way described in Sec. 2 are presented
for four seasons, as well as the corresponding cross-correlation functions
LNCF($\tau$) calculated according to Eq. 3.

\section {Testing the method: processing the mock light curves}

The goal of the TDC campaign  was to let the
researches check their capability to quickly and adequately process and
interpret a huge flow of the observational data expected from the Dark
Energy Survey, PanSTARRS, and LSST, when they become operational.
We used the TDC0 synthetic light curves to test our
method for precision, bias and robustness.

At the first stage of the competition, TDC0, 56 pairs of time series have
been issued, from the most simple cases -- low-noise, well-sampled time
series, without seasonal gaps and microlensing, -- and those ones poorly
sampled, with large gaps and distorted by the effects of noise and
microlensing events.  The simulation team proposed four metrics (Liao et
al. 2015, Dobler et al. 2015), which serve as criteria for the participants
to pass the TDC0 stage. One of them is efficiency, $f$, quantified as a
fraction of light curves, for which the estimates are obtained. Three
other metrics are the goodness of fit quantified by the value of $\chi^2$,
and, according to terminology by Liao et al. (2015), the
"precision" of the estimator $P$ and the "accuracy" or "bias" $A$:
\begin{equation}
\chi^2={1\over{fN}}\sum_j \left({{\Delta t^\prime_j-\Delta t_j}\over
\delta_j}\right)^2,
\end{equation}
\begin{equation}
P={1\over{fN}}\sum_j \left({\delta_j \over {|\Delta t_j|}}\right),
\end{equation}
\begin{equation}
A={1\over{fN}}\sum_j \left({{\Delta t^\prime_j-\Delta t_j}\over
\Delta t_j}\right)
\end{equation}

Here, $N$ is a total number of the proposed light
curves, $\Delta t^\prime_i$ are the time delay values determined by
the participants, with their uncertainties $\delta_j$, and $\Delta
t_j$ are the true time delays.

The simulation team selected the following criteria
to pass the TDC0: $f>0.3$, $0.5<\chi^2< 2$, $P<0.15$, and $A<0.15$.
We were not cautious enough in selecting the results for submission,
having included some ambiguous measurements of the time delays, or
the measurements with too large values of the estimated uncertainties.
This resulted in $f=0.64$ and, quite naturally, in inadmissibly large
values of $P$ and $A$. Meanwhile, rejection of even a few uncertain
results leads to the abrupt changes in values of the metrics. In the
further analysis presented below, we addressed our results
corresponding to $f=0.54$, (a number of successful
determinations is 30). It means that, as compared to the results
submitted to the TDC0 "evil team", we excluded from the present
analysis six determinations for which the individual relative errors
$P^\prime_j={\delta_j/\Delta t^ \prime_j}$ more than three times
exceeded the average relative error $P^\prime$ for the sample under
investigation:
\begin{equation}
P^\prime={1\over{fN}}{\sum_j \left({\delta_j \over|\Delta t^\prime_j|}\right)}\,.
\end{equation}
Therefore, we rejected the uncertain determinations blindly, that is,
we did not address the true time delays, but used the quantities
$P^\prime_j={\delta_j/\Delta t^\prime_j}$ and (7); the latter is called
the "blind" precision in Bonvin et al. (2016).
 \begin{table}
 \centering
 \caption{The metrics A, P and $\chi^2$ (Eqs.4 - 6) calculated for three
 different ranges of the time delays: the short delays,
 $\Delta t<10$,  medium, $10<\Delta t<50$, and long ones, $\Delta t>50$.
 In the last column, the metrics for all the three subsets calculated
 together are presented. The last line shows the subsample
 sizes used in estimating the metrics. }
\begin{tabular}{lcccc}
   \hline
Metrics&$\Delta t<10$ & $10<\Delta t<50$ & $\Delta t>50$&All\\
\hline
P & $1.03\pm0.74$ & $0.14\pm0.08$ & $\,\,0.05\pm0.04$&$0.39\pm0.40$ \\
A & $0.24\pm0.17$ & $0.05\pm0.07$ &$-0.04\pm0.06$&$0.09\pm0.11$ \\
$\chi^2 $&$0.38\pm0.27$ & $0.62\pm0.66$ & $\,\,5.75\pm6.22$&$1.91\pm2.57$ \\
\hline
Number&  9 &  13  &  8  &  30 \\
\hline
\end{tabular}
\end{table}
 \begin{table}
 \centering
 \caption{Robustness test fulfilled for three different
 groups of the time delay values corresponding to the
 short, medium
 and long delays (according to the ranges indicated in Table 1). Our
 estimates $\Delta  t^\prime_j$ with the corresponding
 uncertainties  $\delta_j$, their deviations from the true values
 $\Delta t^\prime_j-\Delta t_j$, and the corresponding relative
 deviations, $A_j={{\Delta t^\prime_j-\Delta t_j}\over \Delta t_j}$,
 are shown for three modes of sampling the light curves.}
  \begin{tabular}{lccc}
   \hline \hline
Sampling mode&$\Delta t^\prime_j\pm \delta_j$&$\Delta t^\prime_j-
\Delta t_j$&$A_j$\\
\hline
\multicolumn{4}{c}{True time delay $\Delta t=2.08$ (rung2-pair4)}\\
\hline
Original set &2.02$\pm$ 0.09&-0.06&-0.030\\
Each 5-th point omitted&2.09$\pm$ 0.09&\,\,0.01&\,\,0.005\\
Each 3-rd point omitted&2.05$\pm$ 0.19&-0.03&-0.014\\
  \hline
\multicolumn{4}{c}{True time delay $\Delta t=28.73$ (rung1-pair8)}\\
\hline
Original set &28.48$\pm$ 3.66&-0.25&-0.009\\
Each 5-th point omitted&27.96$\pm$ 4.43&-0.77&-0.027\\
Each 3-rd point omitted&28.89$\pm$ 3.84&\,\,0.16&\,\,0.006\\
\hline
\multicolumn{4}{c}{True time delay $\Delta t=156.78$(rung3-pair6)}\\
\hline
Original set &157.8$\pm$ 13.4&1.02&0.007\\
Each 5-th point omitted&163.4$\pm$ 18.9&6.62&0.041\\
Each 3-rd point omitted&158.4$\pm$ 7.4&1.62&0.010\\
\hline
\end{tabular}
\end{table}

In testing the method in terms of the metrics (4)-(6),
we dealt with the TDC0 set consisting of the light curves of several types,
which differ in their amplitudes, signal-to-noise ratios, sampling rates,
and contributions from microlensing. Also, in preparing our determinations
for submission, we could not but notice that short time delays demonstrate
in general worse precision as compared to the long and medium ones. As
the simulation team commented in the reply to our submission, it
was typical of many of the submissions. Therefore, we analysed the
metrics for the short, medium and long time delays separately.

We calculated the quantities $A$, $P$ and $\chi^2$ determined by expressions
(4-6) for the following subsets of the submitted light curves differing in
ranges of $\Delta t_j$: for $\Delta t_j \leq 10$ (short delays), $10\leq
\Delta t_j\leq 50$ (medium), and $\Delta t_j > 50$, (long delays). The
results are shown in Table 1. For the short time delays, $\Delta t_j\leq 10$,
the values of $A$ and $P$ exceed the boundaries of the TDC0 criteria, while
$\chi^2$ is noticeably less than permitted that can be explained by
overestimation of the uncertainties $\delta_j$. For $10\leq \Delta
t_j\leq 50$, all the three metrics meet the TDC0 criteria. And finally,
for the longest time delays we have $A=-0.04$ and $P=0.05$,
while $\chi^2$ turns out to be too large. This can be explained by the
presence of a single determination in this subsample, which deviates from
the true time delay by the quantity exceeding the estimated error $\delta_j$,
(this is, in particular, "rung3-pair5" light curves: $\Delta t^\prime_j-
\Delta t_j=5.26$, while the estimated $\delta_j=0.9$).

\begin{figure*}
\centering
\resizebox{17.5cm}{!}{\includegraphics{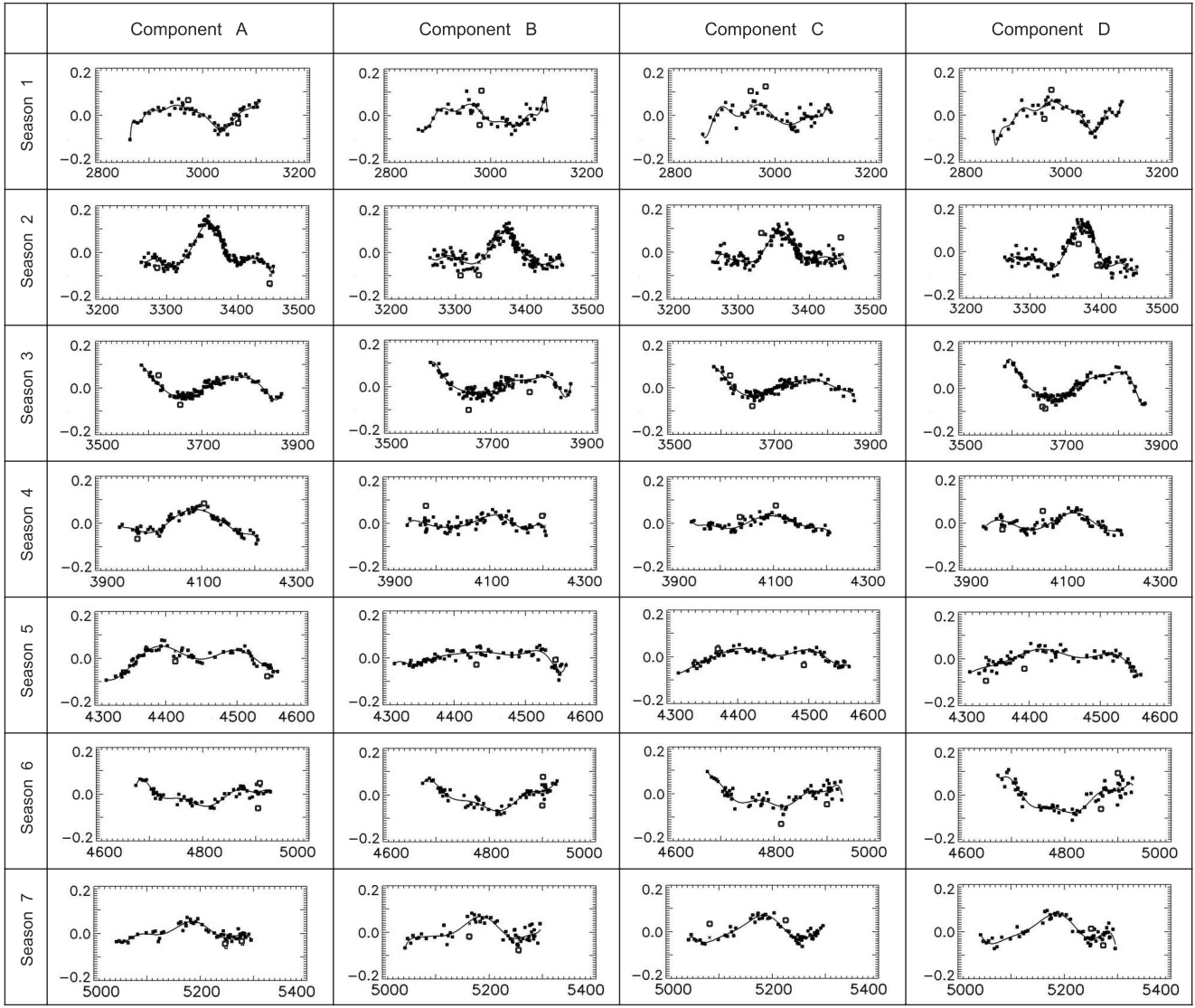}}
 \caption{Light curves of HE 0435-1223 and their approximations by series
 expansions in Legendre  polynomials (image components from A to D are
 from  left to right, seasons 1-7  from top to bottom); the
 low-order terms are excluded. The "bad" points (see in
 the text) are
 marked by squares. The Heliocentric  Julian Dates reduced by 2450000 days
 (HJD-2450000) are along the horizontal  axis.}
\end{figure*}

Table 2 demonstrates some results of testing our method for robustness.
Again, the cases of short, medium and long time delays were
considered separately. The table contains our estimates of the time
delays $\Delta t^\prime_j$ with their errors $\delta_j$ calculated in
the way described in Section 2.2, deviations of our estimates from the
true values, $\Delta t^\prime_j - \Delta t_j$, and the corresponding
relative deviations $A_j=(\Delta t^\prime_j-\Delta t_j)/\Delta t_j$,
(individual signed "accuracies", or "biases").
These values were calculated for three modes of sampling the light
curves: the original data, as they have been proposed for TDC0, the
light curves with each fifth point omitted, and those with each third
point omitted.

Thus, the method is resistant to exclusion of up
to 33\% of the light curve data points, at least for the considered
cases. The degree of robustness is understood to depend on the mode
of the initial data sampling, and on the signal-to-noise ratio
inherent in the initial data photometry.

Closing consideration of the metrics for our estimator,
we would like to note that in 6 cases of 30 successful determinations,
the relative deviations from the true delay values $A_j$ less than 0.01
have been achieved.  As to the failures, their largest
amount took place
for the most difficult (while most realistic) cases of sparse and noisy
data points, with large gaps, as well as
with strong and fast microlensing events, ("rung5" and "rung6" sets).
In Fig. 2 six different types of the simulated light curves are shown
together with our determinations of the time delays
$\Delta t^\prime_j$ and their uncertainties $\delta_j$,
true time delays $\Delta t_j$, and the corresponding relative deviations
of our estimates from the true values $A_j=(\Delta t^\prime_j - \Delta
t_j)/\Delta t_j$. We see two cases of very successful hits here, with
$A_j < 0.01$, as well as one of the examples of our failures ("rung6-pair8"
curves).

\section {Quadruple lens HE 0432-1223}

We applied our method to the excellent data of the 7-years monitoring of
the quadruple gravitation lens HE 0435-1223 (the photometry is available
from the CDS archive, http://cdsarc.u-strasbg.fr/), which have been
used by Courbin et al. (2011) to determine the time delays.

 \begin{table*}
 \centering
 \caption{The estimates of the time delays $\Delta t$
  (in days) for all the six component pairs of the
  HE 0435-1223 quadruple system,  as measured by Kochanek et al.(2006)
  and Courbin et al. (2011), (the data are  taken from the  latter work). }
  \begin{tabular}{llcccccc}
   \hline \hline
Data&Author&$\Delta t_{AB}$&$\Delta t_{AC}$&$\Delta t_{AD}$&
$\Delta t_{BC}$&$\Delta t_{BD}$& $\Delta t_{CD}$\\
\hline
SMARTS(seasons 1,2)&Kochanek(2006)&-8.0$\pm 0.8$&-2.1$\pm 0.8$
&-14.4$\pm 0.8$&&&\\
SMARTS(seasons 1,2)&Courbin(2011), MD &-8.8$\pm 2.4$&-2.0$\pm2.7$
&-14.7$\pm 2.0$&6.8$\pm 2.7$&-5.9$\pm 1.7$&-12.7$\pm 2.5$\\
COSMOGRAIL,7 seasons&Courbin(2011), MD&-8.4$\pm 2.1$&-0.6$\pm 2.3$
&-14.9$\pm 2.1$&7.8$\pm 0.8$&-6.5$\pm 0.7$&-14.3$\pm 0.8$\\
  \hline
\end{tabular}
\end{table*}
\begin{table*}
\centering
\caption{The time delay values (in days) for HE 0435-1223 as determined
 from  different  seasons with the proposed method. For each time delay
 the value of the corresponding cross-correlation function in its
 maximum is indicated in brackets. }
  \begin{tabular}  {l*{6}{c}}
   \hline \hline
Season & $\qquad$ $\Delta t_{AB}$ $\quad$ & $\quad$ $\Delta t_{AC}$
$\qquad$ &
$\quad$ $\Delta t_{AD}$ $\qquad$ & $\quad$ $\Delta t_{BC}$ $\qquad$ &
$\quad$ $\Delta t_{BD}$ $\qquad$ & $\quad$ $\Delta t_{CD}$ $\qquad$\\
\hline
Season 2 (2005) & -11.0 (0.974) & -0.8 (0.988)&-14.5 (0.983)&10.4 (0.995)
&-3.7 (0.989)&-14.7 (0.996)\\
Season 3 (2006)&-10.2 (0.967)&-2.3 (0.990)&-11.9 (0.974)&7.2 (0.968)
&-3.3 (0.978)& -9.5 (0.961)\\
Season 4 (2007)&-10.9 (0.807)&-3.8 (0.934)&-13.1 (0.931)&3.4 (0.891)
&-9.1 (0.893)&-10.4 (0.946)\\
Season 5 (2008)&-10.6 (0.556)&-5.0 (0.963)&-21.3 (0.871)&7.2 (0.603)
&-9.4 (0.771)&-15.3 (0.915)\\
Season 6 (2009)&-10.0 (0.980)&-2.7 (0.957)&-5.8  (0.950)&5.2 (0.942)
&-3.4 (0.956)&-1.6 (0.943)\\
Average &$-10.3\pm 0.4$&$-2.9\pm 1.4$&$-13.3\pm 3.7$&$6.7\pm 2.3$&
$-5.8\pm 2.8$&$-10.3\pm 3.5$\\
\hline
Season 1 (2004)& -1.6 (0.963)& 1.4 (0.829)&-18.5 (0.965)&-0.4 (0.840)
&-22.1 (0.915)&-20.7 (0.910)\\
Season 7 (2010)& -1.6 (0.966)& 0.0 (0.954)& -1.4 (0.951)& 1.3 (0.977)
&2.9 (0.949)& -0.2 (0.964)\\
  \hline
\end{tabular}
\end{table*}
The lens HE 0435-1223 is an example of the systems where microlensing
events can be considered as "slow" ones. The object with a redshift of
$z=1.689$ was identified as a quasar lensed by a $z=0.4546$ galaxy by
Wisotski et al. (2002). The first time delay estimates have been made
by Kochanek et al.(2006) from the two-years monitoring
data obtained in
2004-2005 with the SMARTS 1.3-m telescope. Later on, these data were
supplemented by observations at other telescopes, and observations in
2006-2010 have been added (Courbin et al. 2011). The estimates of the
time delays obtained in these two works are shown in Table 3 taken from
Courbin et al. (2011).

To describe the intrinsic variability of a source quasar, Kochanek
et al. (2006) approximated the data by series expansion
in Legendre polynomials, with the A component selected as the reference
one. The maximum permissible order of the polynomial was determined
with the use of the F-test and constituted 20. The microlensing
brightness variations in individual images with respect to image A
(differential microlensing) were described separately by Legendre
series as well, but of much lower orders - 3 or less. Thus, the model
light curve was composed of two constituents: the quasar light curve
approximated as a Legendre series of the order up to 20, and
microlensing variations represented as a lower-order Legendre series.
Then the problem of searching the parameters is solved, which would
provide the minimal RMS difference between the observed light curve
and the reference (model) one built in the way described above. The
data of each season were processed separately. The results taken with
this approach are presented in Table 3 (the first line).

The second and third lines in Table 3 contain the time delays of
HE 0435-1223 obtained by Courbin et al. (2011) from the results of
the detailed seven-years monitoring. The approach used in their
work is based on the minimum dispersion (MD) method proposed by
Pelt et al. (1996). To avoid the effect of the reference curve
selection, each pair was processed twice, with the change of the
reference component. Then the total dispersion was minimized by
varying the time lags and
parameters of the polynomials
representing microlensing light curves. According to Courbin et
al. (2011), variations of the quasar brightness were least of
all distorted by microlensing in component B.

The light curves of the A, B, C and D image components and the
corresponding regression curves calculated according to the
algorithm described in Sec. 2 can be seen in Fig. 3. Two low-order
terms (the mean level and linear trend) were excluded to represent
the source quasar light curves in seasons 1 to 3 and 6 to 7, while
for seasons 4 and 5 the next (quadratic) term was also needed to
represent microlensing, therefore, three lower-order terms were
excluded from the light curves in Fig. 3 for these seasons.

The orders of polynomials were 17 for the 2-nd season, 15 for
the 4-th and 5-th ones, and 11 for the rest. As is noted in
Section 2, in selecting the polynomial order, we were guided, first,
by a behavior of the RMS error of approximation in its dependence
on the polynomial order, and second, we tried to avoid oscillations
emerging sometimes for too high orders. An increase of the polynomial
order above 17 was found to have a minor effect on the precision of
approximation while producing unacceptable
oscillations at the realization borders. It should be noted that
season 2 is characterized, as compared to others, by the most fast
and high-amplitude variations of flow, thus providing the highest
reliability of the time delay estimates.

The next step was the calculation of the Locally Normalized
Correlation Functions (LNCF) for the pairs of approximating polynomials
exempted from the lower-order terms representing the "slow" microlensing.
The examples of such cross-correlation functions calculated for all the
six permutations of four components in two are presented in Fig. 4 (the
data of the second and fifth seasons were used as the examples of the
"best" and "worst" seasons in the sense of their cross-correlation
maxima values).

The estimates of the time delays for all seasons are shown in Table 4,
together with the results of their averaging over the seasons. The values
of the corresponding cross-correlation functions in their maxima are
shown in brackets. The errors indicated in Table 4 in
the "Average" line
were calculated as the RMS deviations of the values obtained for each
season from the value averaged over all seasons (see Sec. 2.2 for more
explanations).

\begin{table*}
 \centering
 \caption{Values of the time delays in PG1115+080 (in days) from observations
  by Schechter et al., presented in Schechter et al. (1997), Barkana 1997,
  Eulaers et al. (2011), and from observations at the Maidanak Observatory
  in 2004-2006 (Vakulik et al. (2009). In the last two lines, the results of
  re-analysis fulfilled in this work are presented .}
  \begin{tabular}{lcccl}
   \hline \hline
Authors    & $\qquad$  $\Delta t_{BA}$ $\quad$ & $\qquad$ $\Delta t_{CA}$
$\quad$ & $\qquad$ $\Delta t_{BC}$ $\qquad$  &Comments\\
\hline
Schechter et al.(1997) &14.3$\pm$3.4 &9.4$\pm$3.4 & 23.7$\pm$3.4&Method
by Press et al. (1992)\\
Barkana (1997)         &11.7$\pm$2.0 &11.0$\pm$0.9& 25.0$\pm$3.6&Method by
Press et al. (1992)\\
Eulaers \& Magain(2011)& 5.8         &15          & 20.8       &Numerical
Model Fit (NMF)\\
Eulaers \& Magain(2011)& 10.3        &7.6$\pm$3.9 &17.9$\pm$6.9&Minimum
Dispersion Method ((MD)\\
Vakulik et al.(2009)   & 4.4$\pm$3.2 &12.0$\pm$2.4&16.4$\pm$3.4&Joint
estimate over three seasons\\
Vakulik et al.(2009)  &5.0           & 9.4        & 14.4     &For the first
season\\
This work             & 12.9         & 4.2        & 16.8     &For the first
season\\
This work             &9.7$\pm$2.5   & 7.5$\pm$5.3&17.6$\pm$6.9&Averaged over
three seasons\\
\hline
\end{tabular}
\end{table*}

First of all, a very good convergence of estimates for the AB pair between
the seasons should be noted, with the uncertainty as small as 0.4 days,
though the time delay values themselves are systematically larger than
those in (Kochanek et al. 2006) and (Courbin et al. 2011). Somewhat larger
uncertainty, though smaller than that in Courbin et al. (2011), is observed
for the AC pair, with the time delay value consistent with those in Table 3.
The rest of the pairs demonstrate noticeably larger scatter between the
seasons, while the average estimates are rather well consistent with the
results reported by Kochanek et al. (2006) and Courbin et al. (2011),
excluding, perhaps, the CD pair. And finally, a very poor consistence of
the results for seasons 1 and 7 with those for other seasons should be
noted for all the component pairs. The reason may be a somewhat worse
photometry and/or more sparse sampling of the initial light curves,
as compared to other seasons. Surprisingly, it is not always confirmed by
the values of the cross-correlation maxima indicated in the brackets.
\begin{figure}
\resizebox{8.5cm}{!}{\includegraphics{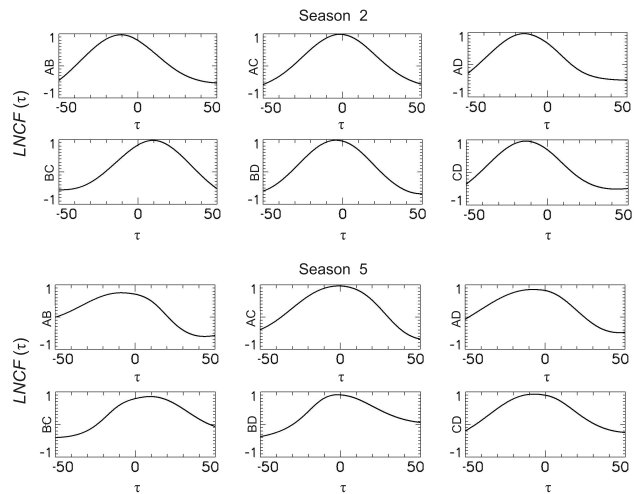}}
 \caption{Locally normalized cross-correlation functions $LNCF(\tau)$
 for pairs of the HE 0435-1223  light curves represented by the regressions
 shown in Fig.3 (seasons 2 and 5).}
\end{figure}

Courbin et al. (2011) report that they tested their curve-shifting method for
stability in different ways, in particular, by processing separate seasons
and groups of seasons. In doing so, they found out no effect on the result
that would be of any significance (unfortunately, the time delay values for
individual seasons are not presented in their paper). They also note that
stability of estimations will be much worse if the data for only two or three
seasons are used, and stress the importance of many years of monitoring at a
good sampling rate.

Therefore, the time delays presented in Table 4 in the
line marked as "Average" are consistent with the results obtained by
Kochanek et al.(2006) and Courbin et al. (2011), -- at any rate, within
the error bars, which have been estimated in these works with the method
of statistical trials, as is generally accepted.

\begin{figure*}
\resizebox{17.4cm}{!}{\includegraphics{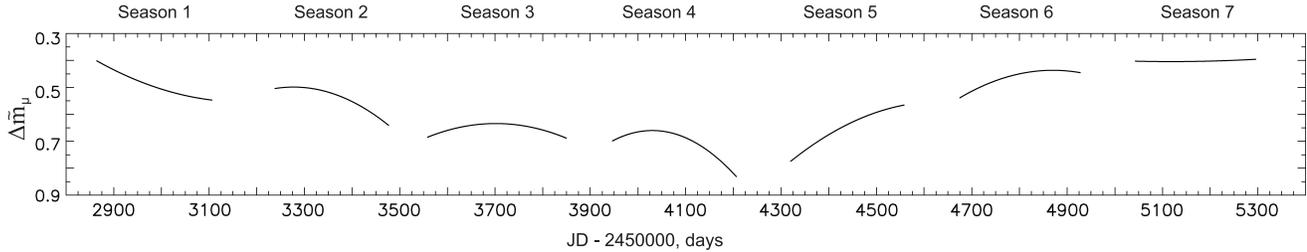}}
 \caption{Differential microlensing light curves of the A component with
 respect to B represented as differences of low-order terms of the
 corresponding polynomial regressions. The heliocentric Julian dates
 minus 2400000 are along the horizontal axis, and the magnitude difference
 {$\mathrm{\Delta\tilde{m}_{\mu}}$} is along the vertical axis.}
\end{figure*}

In conclusion, we would like to note that our approach provides a very
simple way to display a contribution from microlensing events to the
observed quasar variability for each season separately. To do this, one
must just restore the low-order terms,
which have been eliminated earlier from the polynomials approximating
the observed light curves, as is described above. These low-order terms
contain both the microlensing and quasar variability constituents. Since
the quasar variability are the same in different quasar images, the
difference between these low-order polynomials with the time shifts
equal to the corresponding time delays, will describe differential
microlensing. In Fig. 5, such differential microlensing light curves
for the A-B pair are presented for seven seasons, which are clearly seen
to be consistent with the similar curves in fig. 4 by Courbin et al.
(2011) calculated in another way.

\begin{figure*}
\resizebox{15cm}{!}{\includegraphics{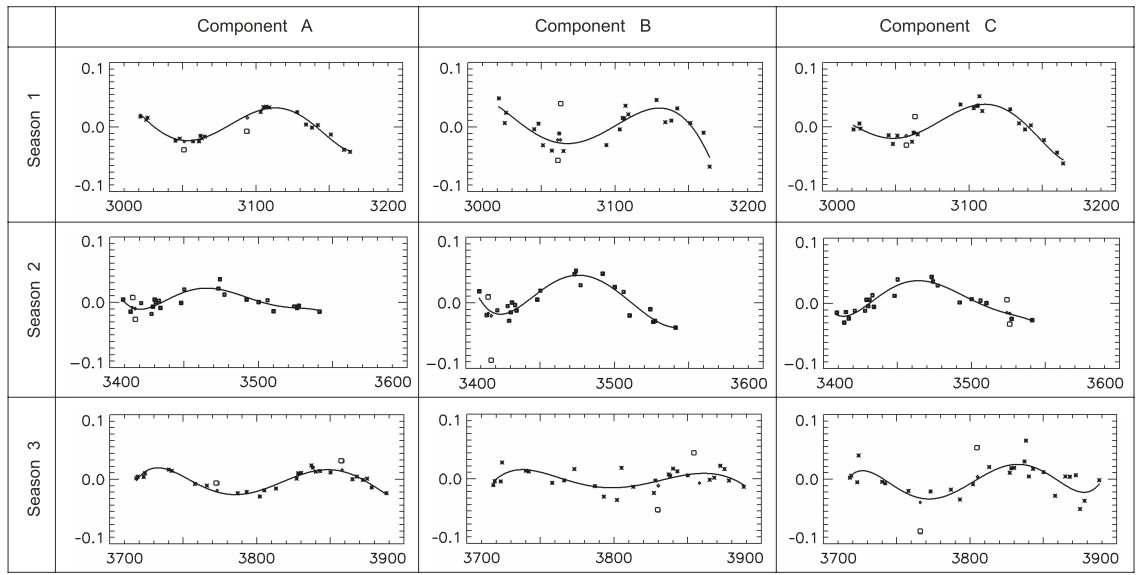}}
 \caption{Light curves of the A, B and C components of PG1115+080 from
 observations in 2004, 2005 and 2006 (seasons 1-3), and their
 approximations by series expansions in Legendre polynomials up to the
 5-th order without the  zero and first-order terms. The "bad" points
 (see Sec. 2) are shown by squares. }
\end{figure*}

\section {Quadruple lens PG 1115+080}

The quadruply lensed quasar PG1115+080 is among the systems, which need
more accurate estimates of the time delays. This is the second gravitational
lens system, for which the time delays have been measured (Schechter et al.
1997). Soon after the first measurements, revision of the same data with
another algorithm was published (Barkana 1997), which in general confirmed
the results by Schechter et al. (1997). New determinations of the time delays
from observations of 2004-2006 at the Maidanak Observatory were reported
in 2009 (Vakulik et al.). The next attempt to reanalyse the data by
Schechter et al. was made by Eulaers \& Magain (2011) with the use of two
methods. All the known time delay estimates for PG 1115+080 are collected
in Table 5.

The data presented in Table 5 explain our desire to reprocess the data by
Vakulik et al.(2009) with the use of our new algorithm. Indeed, while the
time delay estimates for pairs BA and AC deviate from each other in
different authors rather randomly, the estimates of $\Delta t_{BC}$ in
Vakulik et al. (2009) and Eulaers \& Magain (2011) are definitely lower
than those presented by Schechter (1997) and Barkana (1997). Note, in
particular, a good agreement between $\Delta t_{BC}$=17.9 days obtained
by Eulaers \& Magain (2011) with the Minimum Dispersion method,
and $\Delta t_{BC}$=16.4 days in Vakulik et al. (2009).

In Fig. 6, three light curves for three seasons are shown together with
the corresponding approximations by series expansions in the normalized
Legendre polynomials. Because of the smallness of the time delays between
A1 and A2 predicted by the macrolens model from the
system geometry,
their light curves were joined to form a single A light curve.

Similar to the light curves of HE 0435-1223 in Fig. 3, the terms of the
zeroth and first orders are omitted in displaying in Fig. 6.
Qualitatively, the mutual time shifts are visible by sight in all seasons,
namely, the B component  leads A and C quite evidently. Concerning the
A-C pair one can say only that the expected time delay value is rather
small. The least scatter of the data points with respect to the
approximating polynomial is observed for the A component, and the largest
one is for B, that is quite natural since the latter is the faintest one.

 The largest time delay value, $\Delta t_{BC}$, which equals 23.7 days
 in Schechter et al. (1997) and 25.0 days in Barkana (1997), was estimated
 by Eulaers \& Magain (2011) to equal 20.8 days in reprocessing the same
 data with the NMF (Numerical  Model Fit) method, and 17.9 days with the
 MD (Minimum Dispersion) method. The latter value is consistent both with
 that obtained from the 2004-2006 data by Vakulik et al. (2009) -- 16.4
 days, and with the results of reprocessing in  the present work --
 16.8 days.

The time delay values for the two other image pairs are consistent in
different works much worse. In this respect, compare a scatter of the time
delay values from Table 5 with the indicated estimates of errors. We see
$\Delta t_{CA}$ varying from 7.5 to 12 days in different authors, with the
error estimates from 0.9 to 3.9 days; $\Delta t_{BC}$ ranges between 16.4
and 25 days, with the errors varying between 3.4 and 6.9 days; and finally,
$\Delta t_{BA}$ varies from 4.4 to 14.3 days, while the errors are from 2.0
to 3.4 days. As is seen, the values of delays vary in a wider range than
it follows from the estimates of their error bars. This fact should not
in effect wonder: investigators know that estimates of the time delays may
be sensitive to the patterns of random fluctuations of sampling points,
which are often indistinguishable from the actual signal features. This
concerns especially the sparse and scanty data, that
is just the case
for the PG 1115+080 data, both by Schechter et al. (1997) and Vakulik
et al. (2009). The results of averaging the estimates of differential
delays obtained in this work from different seasons are shown in the last
line of Table 5, and the results for only the first season are shown in
the last but one line.

It is evident that the time delays in the PG 1115+080 system need to be
further specified, but it is evident also that no essential progress can be
expected from processing the available data. Long-term monitoring with a
sufficient sampling rate is needed, which would provide new high-quality
photometric data.

\section{Discussion and conclusions}

Summarizing, we would like to note the following.
\begin{itemize}
\item  The proposed method implies a pair-wise comparison of light curves
represented by their polynomial approximations. In this respect, our
algorithm is similar to the regression difference technique illustrated
by Tewes et al. (2013) in calculations of the time delays in HE 0435-1223.
\item We tested the method for robustness and bias using
the mock light curves issued for the TDC0 campaign (Liao Kai et al. 2015).
The method demonstrates resistance to exclusion of up to 33\% of light
curve data points and shows no noticeable change in the relative deviations
from the true delay values $A_j$ exceeding 0.04. The testing has also
shown that in 6 cases of 30 successful determinations the value of $A_j$
is less than 0.01.
\item The estimates of the time delays in the quadruply lensed quasars
PG 1115+080 and HE 0435-1223 obtained with the method proposed in this
work are consistent with those obtained for these objects with other
methods earlier. As is shown in Sec. 3, the short time
delays are measured with the worst relative precision (see Table 1).
In PG 1115+080 and HE 0435-1223, we deal with just this case.
\item  The differences of our approach from those proposed by other
authors earlier do not have a fundamental nature, but provide
convenience in calculations. In particular, our approach allows to
exclude or add some of the approximating polynomial terms without a
necessity to recalculate the coefficients. This provides a simple
way to mitigate the effects of microlensing for the
case of "slow" microlensing events. The method can be useful for a
preliminary express analysis of the data flow expected
from the future sky survey programs.
\end{itemize}

 \section*{Acknowledgments}
 The authors thank the Science and Technology Center in Ukraine (STCU
 grant  U127), which had made possible observations at the Maidanak
 Observatory and the further  working on the data. We also appreciate
 the financial support from the Target Program of the National Academy
 of Sciences of Ukraine "CosmoMicroPhysics". The present paper
 has been encouraged by O.I.Bugaenko, for what we appreciate him greatly.

\end{document}